\documentclass[preprint,12pt]{elsarticle}

\usepackage{amssymb}
\usepackage{xcolor}
\usepackage[most]{tcolorbox}
\usepackage{longtable}
\usepackage{enumitem}
\usepackage{amsmath}
\usepackage{booktabs,array}
\usepackage{nicematrix}
\usepackage{dcolumn,longtable,booktabs}
\usepackage[utf8]{inputenc}
\usepackage[symbol]{footmisc}
\usepackage{soul}
\sethlcolor{cyan}

\usepackage{eqnarray} 
\usepackage[numbers]{natbib}
\usepackage{graphicx}
\usepackage{setspace}
\usepackage{amsthm}
\usepackage{dirtytalk}

\usepackage{tikz}
\usetikzlibrary{arrows.meta,positioning}
\usepackage{multicol}
\usepackage{multirow}
\usepackage{algorithm}
\usepackage{algpseudocode}
\usepackage{float}
\usepackage{makecell}

\usepackage[margin=1in]{geometry}

\newlist{legal}{enumerate}{10}
\setlist[legal]{label*=\arabic*.}

\usepackage[breaklinks=true,
  bookmarks,
  pdfpagemode=UseOutlines,
  pdfpagelayout=SinglePage]{hyperref}
\hypersetup{
    colorlinks=true,    
    linkcolor=blue,
    filecolor=magenta,      
    urlcolor=cyan,
}

\definecolor{ztodobg}{RGB}{255,249,227}

\journal{journal}

\begin{document}

\begin{frontmatter}



\title{To Give or Not To Give? Pandemic Vaccine Donation Policy}


\author[Gordon College]{Abraham Holleran}

\affiliation[Gordon College]{organization={Gordon College, Mathematics and Computer Science},
            addressline={255 Grapevine Rd}, 
            city={Wenham},
            postcode={01984}, 
            state={MA},
            country={USA}}

\author[Harvey Mudd College]{Susan E. Martonosi}
\author[Gordon College]{Michael Veatch}

\affiliation[Harvey Mudd College]{organization={Harvey Mudd College, Mathematics},
            addressline={301 Platt Blvd.}, 
            city={Claremont},
            postcode={91711}, 
            state={CA},
            country={USA}}

\begin{abstract}
The global SARS-CoV-2 (COVID-19) pandemic highlighted the challenge of equitable vaccine distribution between high- and low-income countries.  
Many  high-income countries were reluctant or slow to distribute extra doses of the vaccine to lower-income countries via the COVID-19 Vaccines Global Access (COVAX) collaboration \cite{ClintonYoo}.  In addition to moral objections to such vaccine nationalism, vaccine inequity during a pandemic could contribute to the evolution of new variants of the virus and possibly increase total deaths, including in the high-income countries.  
Using the COVID-19 pandemic as a case study, we use the  epidemiological model of Holleran \textit{et al.} \cite{HolleranEtAl_Opt} 
that incorporates virus mutation.  We identify realistic scenarios under which a donor country prefers to donate vaccines before distributing them locally in order to minimize local deaths during a pandemic.  We demonstrate that a nondonor-first vaccination policy can delay, sometimes dramatically, the emergence of more-contagious variants.  Even more surprising, donating all vaccines is sometimes better for the donor country than a sharing policy in which half of the vaccines are donated and half are retained because of the impact donation can have on delaying the emergence of a more contagious virus. Nondonor-first vaccine allocation is optimal in scenarios in which the local health impact of the vaccine is limited or when delaying emergence of a variant is especially valuable.  In all cases, we find that vaccine distribution is not a zero-sum game between donor and nondonor countries.
 Thus, in addition to moral reasons to avoid vaccine nationalism, donor nations can also realize local health benefits from donating vaccines. The insights yielded by this framework can be used to guide equitable vaccine distribution in future pandemics. 
\end{abstract}

\begin{highlights}
\item  A nondonor-first vaccination policy can delay, sometimes dramatically, the emergence of more-contagious variants.  
\item Delaying the variant leaves more time for both donor and nondonor areas to become vaccinated in advance of a second wave, resulting in fewer total deaths. 
\item As a result, donating all vaccines is sometimes better for the donor country than a donor-first or a sharing policy.  
\item Vaccine distribution is not a zero-sum game between donor and nondonor countries: policies other than donor-first can decrease both donor and nondonor deaths. 
\item Nondonor-first vaccine allocation is optimal when the local health impact of the vaccine is limited (highly responsive social distancing measures or a low willingness to be vaccinated in the donor country) or when delaying emergence of a variant is especially valuable (significantly higher infectiousness of the variant).
\item This work highlights that in addition to moral reasons to encourage wealthy nations to share their vaccine stockpile, local public health arguments can also be made in support of vaccine donation.
\end{highlights}


\end{frontmatter}

\section{Introduction} 


In December 2020, the first vaccines against COVID-19 became available in global markets, bringing hope that the global pandemic  that began in late 2019 might be nearing its end \cite{NYT_PF95,moderna}.  The rapid development of these vaccines was funded, in part, by subsidies and contracts provided by governments of several high-income nations, including the United States \cite{pa, ma}.  
High-income nations such as the U.S. secured large stockpiles of COVID-19 vaccines while lower-income nations, particularly those in Africa, relied on the goodwill of these nations to provide access to doses of vaccine \cite{Gavi}.  The World Health Organization (WHO) aimed for all countries in the world to vaccinate 
70\% of their populations by the middle of 2022.  However, many countries, particularly in Africa, were unable to meet these targets, despite the ample supply of vaccines globally \cite{WHO}.  


In addition to moral objections against inequitable vaccine distribution \cite{HunterEtAl_2022_NEJM, AsundiOLearyBhadelia_2021_CHM, Ghebreyesus_2021_PLOS}, there are other reasons why so-called vaccine nationalism is problematic during a pandemic.  First, as was the case with COVID-19, rising cases in one geographic area can trigger surges throughout the world.  Failing to distribute vaccines uniformly to the world's population can serve to prolong a pandemic and its economic disruptions, even in countries able to attain high vaccination rates \cite{WHO}.  Moreover, low vaccination rates and limited health care infrastructure can contribute to the emergence of variants, and such variants can be more contagious and resistant to available vaccines \cite{WSJ, Rockefeller}.


Given the additional ramifications of inequitable vaccine distribution cited above, it is possible that vaccine nationalism could undermine a wealthy nation's own best interests \cite{AsundiOLearyBhadelia_2021_CHM}.  
Using data from the COVID-19 pandemic as a case study, we use an epidemiological disease transmission model 
to examine allocation policies of a wealthy nation's vaccine supply between its own population and that of another geographic regions.
 We compare three policies: donor-first, nondonor-first, 
 and a ``fairness'' policy in which 
 a limited percentage of available daily vaccine doses may be retained by the donor country.  
 Our model identifies realistic scenarios under which a donor country prefers to donate vaccines \textit{before} distributing them locally in order to minimize local deaths.  We demonstrate that a nondonor-first vaccination policy can, under some circumstances, dramatically delay the emergence of more-contagious variants. Thus, there are local public health arguments in favor of vaccine donation, in addition to moral arguments.    
 Although this paper focuses its disease transmission model and parameter estimation on characteristics of COVID-19, the same approach can be applied to other communicable diseases and guide future pandemic policy.


In the next section, we summarize the literature on vaccine allocation policies.  Section \ref{sec:SEIR} describes the epidemiological framework we use to model disease transmission during a pandemic.  
Section \ref{sec:results} presents the results of our approach.  We conclude  in Section \ref{sec:conc}.

\section{Literature Review} 
\label{sec:litreview}

The purpose of this paper is to examine vaccine nationalism in the context of a global pandemic.  Using the COVID-19 pandemic as a case study, we apply the SEIR (Susceptible, Exposed, Infected, Recovered) compartment model developed by Holleran \textit{et al.}.  We direct the reader to the references in \cite{HolleranEtAl_Opt} for context on the use of compartment models for this purpose. To apply their SEIR model to the COVID-19 pandemic, we present parameter estimation and the literature used to support those estimates in  \ref{sec:param}.

We focus this literature review on the question of vaccine allocation between heterogenous populations during a pandemic.  Two papers consider the effectiveness of COVID-19 interventions in the context of population heterogeneity.  Volpert, Vitaly, and Sharma 
examine the effectiveness of vaccination within a heterogeneous population of high transmission and low transmission subpopulations that arise due to characteristics such as age, religious practices, professional experiences, and cultural norms \cite{VolpertVitalySharma_EcoComp_2021}.  They find that 
the effectiveness of vaccination depends on transmission characteristics 
in each group: achieving a high vaccination rate within the high transmission subpopulation leads to lower overall population rates of infection, but a high vaccination rate only within the low transmission subpopulation serves only to protect the low transmission population against infection. Dolbeault and Turinici also consider the impacts of high- and low-transmission subpopulations in the context of lockdown policies in France \cite{DolbeaultTurinici_MathMod_2020}.  The analysis we present examines the interaction between heterogeneous populations (e.g., distinct countries) occurring through the emergence of a variant.  We are thus able to examine contexts in which donating vaccines to low-income countries might reduce COVID-19 deaths in the donor country. 

Work related to the allocation of COVID-19 vaccines generally has a localized focus.  The work of Bicher \textit{et al.} examines how to prioritize vaccine recipients within a nation based on factors such as age and vulnerability \cite{BicherEtAl_PLOS_2022}. Pan \textit{et al.} examine vaccine allocation to public and private hospitals responsible for distributing the COVID-19 vaccine.  They model the role of information-sharing and subsidies in incentivizing private sector participation in vaccine distribution and maximizing vaccine uptake \cite{PanEtAl_AOR_2022}. Tavana \textit{et al.} develop a mixed integer linear programming model for operational and tactical decisions surrounding COVID-19 vaccine distribution in developing countries that accounts for vulnerable subpopulations and distribution complexities such as the availability of cold-chain infrastructure \cite{TavanaEtAl_AOR_2021}.  Their model determines which types of vaccines should be procured, given storage and distributional considerations, and where distribution centers should be located, under the assumption that the country already has a mechanism for obtaining the vaccine. Van Oorschot \textit{et al.} examine the distribution of COVID-19 diagnostic tests, comparing COVID-19 transmission in Norway under an isolation policy to that under one-sided donation during times of surplus, one-sided receipt during times of shortage, and two-sided donation of excess and receipt of shortage \cite{VanOorschotEtAl_2022}.  However, they do not consider how the disease transmission dynamics in partner countries affect disease transmission in the country of consideration.

In the case of the COVID-19 pandemic, transmission has occurred even among geographically distant countries, and variants have rapidly spread from one geographic area to another. The work of Duijzer \textit{et al.} predates the COVID-19 pandemic and examines allocation of influenza vaccine stockpile to multiple subpopulations, in both non-interacting and interacting scenarios, using an SIR model \cite{DuijzerEtAl_2018_POM}.  For the non-interacting scenario, reflecting geographically distant populations, they find that heavily vaccinating certain subpopulations while leaving other subpopulations unvaccinated maximizes total health benefit but contributes to health disparities.  As interaction increases between the subpopulations, Duijzer \textit{et al.} find that health disparities under 
the optimal vaccine allocation policy persist but diminish.  

Rotesi \textit{et al.} examine conditions under which a nation should donate vaccines \cite{RotesiEtAl2021}.  They formulate an SIR epidemiological model that incorporates travel between different countries, and they simulate the impact of different vaccine donation policies.  They demonstrate that it is beneficial to donate vaccines when the donor and recipient countries are close to the herd immunity limit. Because COVID-19 cases dramatically increase just below the herd immunity limit, donating vaccines to prevent the reintroduction of COVID-19 from outside countries is more beneficial near that limit. 

This paper examines a similar question. Using the epidemiological model of \cite{HolleranEtAl_Opt}, we simulate several vaccine allocation policies using parameters estimated from COVID-19 data (see \ref{sec:param}).  Although this model does not directly capture travel between geographic regions, it incorporates the emergence of more  contagious variants that appear in the donor country after a time lag.  Thus, this paper sheds light on benefits for vaccine-sharing between countries in the context of geographic interaction and rapid mutation.


\section{SEIR Model} 
\label{sec:SEIR}

We adopt the Susceptible, Exposed, Infectious, and Recovered model with additional states for vaccinated individuals (SEIR-V) of Holleran \textit{et al.} \cite{HolleranEtAl_Opt}. The reader is directed to that paper for the details of the model, its derivation, and analysis.  The model multiple geographic areas $a \in \mathcal{A}$ that interact through virus mutation.  For the purposes of this paper, we consider two areas consisting of a donor region and a nondonor region.  A more contagious variant emerges in the nondonor area after a given number of infections and then spreads to the other areas after a fixed time lag. Unlike in \cite{HolleranEtAl_Opt}, we assume here a deterministic emergence of the variant for simplicity.  We now restate key elements of the model. 

\subsection{SEIR Model with Vaccination}
\label{ss:SEIRwVax}

\begin{figure}[!ht]
    \centering
\begin{tikzpicture}[node distance=1cm, auto,
    >=Latex, 
    every node/.append style={align=center},
    int/.style={draw, minimum size=1cm, circle}]
   \node [int] (S)             {$S$};
   \node [int, right=of S] (E) {$E$};
   \node [int, right=of E] (I) {$I$};
   \node [int, right=of I] (D) {$D$};
   \node [int, above=of D] (R) {$R$};
    \node [int, above=of S] (SV) {$S^V$};
    \node [int, above=of E] (EV) {$E^V$};
    \node [int, above=of I] (IV) {$I^V$};

   \path[->] (S) edge node {} (E)
   (E) edge node {} (I)
   (I) edge node {} (D)
   (I) edge node {} (R)
   (S) edge node {} (SV)
   (SV) edge node {} (EV)
   (EV) edge node {} (IV)
   (IV) edge node {} (D)
   (IV) edge node {} (R);
\end{tikzpicture}
    \caption{State diagram for the single-area SEIR-V model of \cite{HolleranEtAl_Opt}}
    \label{fig:SEIR-V}
\end{figure}
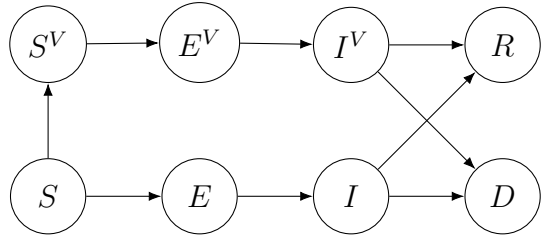

The states of the model are Susceptible ($S$), Exposed ($E$), Infectious ($I$), 
Dead ($D$), and Recovered ($R$), plus $S^V$, $E^V$, and $I^V$ if also vaccinated (Figure \ref{fig:SEIR-V}). All quantities that depend on area have the subscript $a$; however, it will be suppressed whenever possible for simplicity. For example, $S_a(t)$ denotes the number of people in state $S$ in area $a$ at time $t$, but we often refer to it simply as $S$ or $S(t)$ when referring to a generic region.  

As in \cite{HolleranEtAl_Opt}, the model assumes that: (1) Vaccinated individuals have lower rates of becoming infected, lower mortality, and are less contagious. (2) The donor area has a daily budget of $B$ vaccine doses to distribute to itself and other areas. The distribution policy is given by $V_a(t)$, the number of doses in each area $a$ on day $t$. (3) Only a proportion $\rho_a$ of the population in area $a$ is willing to be vaccinated.  Letting $W_a(t)$ denote the number of individuals in state $S$ in area $a$ at time $t$ who are willing to be vaccinated, vaccinations stop in area $a$ once $W_a(t)=0$. (4) People are more cautious when there is a surge in cases, so the effective number of infectious individuals is attenuated when infections are high (see Eq. (\ref{eq:vectors})). (5)  The new variant has a higher infection rate than the previous variant.  Unlike in \cite{HolleranEtAl_Opt}, we assume a deterministic emergence of the new variant, which is discussed below.  (6) The time horizon of the model, $T$, is sufficiently short (180 days, or approximately 6 months) so that individuals cannot be infected twice, and only one new variant will emerge. (7) The model allows daily reallocation of vaccines without logistical constraints.

The parameters of the model are the same as in \cite{HolleranEtAl_Opt} and are listed in Tables \ref{tab:param1} and \ref{tab:param2} for reference. More discussion of their estimated values is given in \ref{sec:param}. All parameters are assumed to be constant over time except for $V_a(t)$, which is the vaccine allocation policy in each area, and $a(t)$, the infection rate, which changes due to mutation.

{\small \begin{longtable}{| p{0.15\textwidth} | p{0.18\textwidth} | p{0.2\textwidth} |p{0.45\textwidth}|}
\caption{Parameters of SEIR-V that depend on the area \cite{HolleranEtAl_Opt}}
\label{tab:param1}
    \\ \hline
    Parameter & Base Value & Sensitivity Analysis & Description  \\ \hline
    $N_a$ & 100,000 &$-$ &Initial population  \\ \hline
    $\rho_a$ & 0.78 & [0.5, 1.0] &Proportion of population willing to be vaccinated  \\ \hline
    $\rho_a^V$ & 0 & $-$  &Initial proportion of population vaccinated \\ \hline
    $\rho_a^I$   & Donor: $3.6 \times 10^{-3}$ Nondonor: $7.2 \times 10^{-4}$ & $[2.6 \times 10^{-3}, 4.9 \times 10^{-3}]$ & Initial new cases per day  as a proportion of the population \\ \hline
    $V_a(t)$ & Policy-dependent & $-$& Rate of vaccinations available at time $t$ (people/day)  \\ \hline
    $r_0$ & $1/3.9$ & $[1/5.1, 1/2.5]$ & Rate out of the infectious state without testing
    (proportion/day)  \\ \hline
    $\Delta r_a$ & 0.035 & [0.026, 0.060] & Contribution of testing to the rate out of the infectious state in the donor area (proportion/day)  \\ \hline
    $r_a^d$ & $r_a^d = r_0 + \Delta r_a$. &$-$ & Rate out of the infectious state (proportion/day)  \\ \hline
    $\gamma_a$ & $1$ & $2$ (nondonor)  &Infection multiplier for an area \\ \hline
    $\alpha_a(t)$ & Computed. 
    & $-$ & Infection rate of out of the susceptible state (proportion/day)  \\ \hline
 \end{longtable}}

{\small 
\begin{longtable}{| p{0.15\textwidth} | p{0.18\textwidth} | p{0.2\textwidth} |p{0.45\textwidth}|}
\caption{Other parameters of SEIR-V \cite{HolleranEtAl_Opt}}
\label{tab:param2}
    \\ \hline
    Parameter & Base Value  & Sensitivity Analysis & Description \\ \hline
    $r^I$ & 1/5 &$-$&Rate out of the exposed state (proportion/day) \\ \hline
    $p^D$ & 0.014 &$-$&Unvaccinated mortality rate \\ \hline 
    $p_V^D$ & 0.0079 &$-$& Vaccinated mortality rate  \\ \hline 
    $a_0$ & 0.6 &$-$& Initial infection rate (proportion/day). 
    \\ \hline
    $\Delta a$ & 0.6 & [0.3, 0.9] & Change in infection rate for a new variant (proportion/day). 
    \\ \hline
    $v^u$ & 0.03 &[0.025, 0.10]&  Upper limit on proportion of population infectious, due to behavioral changes. 
    \\ \hline
    $p^e$ & 0.6 &$[0.5, 0.8]$ & Transmission rate from a vaccinated person as a proportion of rate for an unvaccinated person; $1 - p^e$ is the vaccine effectiveness against transmitting the virus  \\ \hline
    $p^r$ & 0.6 &$[0.5, 0.8]$& Infection rate for a vaccinated person as a proportion of rate for an unvaccinated person\\ \hline
    $n$ & 45,000 & [60,000, 90,000] & Person-days in the infectious state before new variant appears. Only the nondonor area and unvaccinated individuals are counted. \\ \hline
    $L$ & 15 &$-$& Lag for the variant to reach other areas (days) \\ \hline
    $T_D$ & 25 &$-$& Time for a variant to dominate, i.e., represent half the new cases in an area (days) \\ \hline
    $p$ & 0.01 &$-$& Proportion of people in state $I$ and $I^V$ that have the new variant when it is introduced in an area\\ \hline
    $k$ & $k = \ln[(1-p)/p]/T_D$. 
    &$-$&  Rate parameter for when the new variant dominates\\ \hline
    $B$ & 1500 & $[1000,2000]$ & Daily budget of vaccine doses available to be distributed by the donor area\\ \hline
    $T$ & 180 & $[270,360]$ & Simulation time horizon (days) \\ \hline
 \end{longtable}
 }

The system of differential equations for the SEIR-V model \cite{HolleranEtAl_Opt} is
\begin{align}
    \frac{\text{d}S}{\text{d}t}&=  -V^*(t) - \alpha(t) \frac{S(t)}{N}\mathcal{V}(t) \nonumber\\
    \frac{\text{d}S^V}{\text{d}t}&=  V^*(t) - p^r\alpha(t) \frac{S^V(t)}{N}\mathcal{V}(t) \nonumber \\
    \frac{\text{d}E}{\text{d}t} &= \alpha(t) \frac{S(t)}{N}\mathcal{V}(t) -r^IE(t) \nonumber  \\
    \frac{\text{d}E^V}{\text{d}t} &= p^r \alpha(t) \frac{S^V(t)}{N}\mathcal{V}(t) -r^IE^V(t) \nonumber  \\
    \frac{\text{d}I}{\text{d}t}&= r^IE(t) - r^dI(t)                  \label{eq:SEIRVTM} \\
    \frac{\text{d}I^V}{\text{d}t}&= r^IE^V(t) - r^dI^V(t) \nonumber \\   
    \frac{\text{d}D}{\text{d}t} &= r^dp^DI(t) + r^dp_V^DI^V(t) \nonumber  \\
    \frac{\text{d}R}{\text{d}t} &= r^d(1-p^D)I(t) + r^d(1-p^D_V)I^V(t), \nonumber
\end{align}
where $\mathcal{V}(t)$, $V^{*}(t)$, and $\alpha(t)$  are defined as follows:

\textbf{Vaccine- and Behavior-Adjusted Infectious, Non-Isolated People \cite{HolleranEtAl_Opt}:}
\begin{equation}
    \mathcal{V}(t) = \left( 1 - \frac{I(t) + p^eI^V(t)}{Nv^u} \right)
    [I(t) + p^eI^V(t)]. \label{eq:vectors}
\end{equation}

\textbf{Vaccinations Administered \cite{HolleranEtAl_Opt}:}
\begin{equation}
    V^{*}(t) = \begin{cases}
        V(t) & \text{for } t \text{ before } W(t)=0 \text{ in a given area}\\
        0 & \text{for } t \text{ after } W(t)=0 \text{ in a given area}
    \end{cases}\label{eq:vaxadmin}
\end{equation}

\textbf{Emergence of Variant:}

Holleran \textit{et al.} use a stochastic model for the emergence of a more infectious variant \cite{HolleranEtAl_Opt}.  Here, we propose a simpler, deterministic model of how the infection rate changes over time due to a new variant, the details of which can be found in \ref{sec:emerge}. Letting $a_0$ be the initial infection rate of the virus, $\Delta a$ be the increase in infectiousness of the new variant, area $m$ be the area in which the new variant first emerges (assumed to be the nondonor area in the baseline scenario), $t_n$ be the day on which the new variant appears (chosen as the time until $n$ unvaccinated infectious person-days have accumulated.), $T_D$ be the delay until the new variant becomes dominant (half of infections), $L$ be the time lag until the variant reaches the other area,  $k = \ln[(1-p)/p]/T_D$ be the rate parameter, and $\gamma_a$ be the behavior factor for area $a$, the time-varying infection rate is given by
\begin{equation} \label{eq:alpha_a}
  \begin{aligned}
    \alpha_m(t) & = a(t) \gamma_m , \quad t \ge t_n  \\
    \alpha_a(t) & = \alpha_m(\, \max\{t-L,\,0\} \, )\gamma_a/\gamma_m \text{   for }a \neq m,
  \end{aligned}
\end{equation}
where 
\begin{equation}\label{eq:alpha}
    a(t) = a_0 + \frac{\Delta a}{1+e^{-k(t-(t_n + T_D))}}.
\end{equation}

\section{Results}
\label{sec:results}

The following scenarios are referenced, all of which have two areas (donor and nondonor):
\begin{itemize}
    \item \textit{Baseline:} This scenario uses the base values in Tables \ref{tab:param1}-\ref{tab:param2}, and a duration of $T = 180$ days. 
    However, nondonor initial cases are reduced to 20\% of donor initial cases, i.e., $\rho^I = 7.2 \times 10^{-4}$ for nondonor.  Only nondonor area infections contribute to the emergence of a variant.
    \item \textit{Large nondonor initial cases:} From the baseline scenario, nondonor initial cases are reset to 100\% of donor initial cases and infectious days before the variant is $n$ = 60,000.
    \item \textit{Global variant:} The baseline scenario, but the variant emerges after $n=45,000$ or $n=90,000$ infectious days in all areas (instead of just the nondonor area).
   \item \textit{No variant:} The baseline scenario, but without the emergence of the variant.
      \item \textit{No vaccine:} The baseline scenario, but without vaccinations. 
\end{itemize}

We simulate three policies: donor-first, nondonor-first, and a “fairness” policy. 
The fairness policy limits the percentage of daily vaccine doses that the donor country is allowed to retain, ensuring that the remaining doses are donated in parallel. 

\subsection{Emergence of the Variant for Different Policies and Scenarios}
\label{sec: emergence}

Table \ref{tab:deaths} summarizes the results of the different policies and scenarios, as measured by donor-country deaths, total deaths, and date of variant emergence.  Unsurprisingly, deaths in the donor country and in total are highest in the scenario of no vaccine.  Deaths are lowest, regardless of vaccine allocation policy, in the no variant scenario.  Moreover, when there is no risk of a more-contagious variant emerging, the donor-first policy is optimal,  with respect to both donor country deaths and total deaths.

\begin{table}[!h] 
\caption{Effect of policy, vaccination, and scenario}
\label{tab:deaths}
\center
\begin{tabular}{| l | l | c c | c |}
\hline 
&& \multicolumn{2}{c|}{Deaths } & Time of \\
Scenario & Policy & Donor & Total & variant (days) \\ \hline \hline
No vaccine & n/a & 965 & 1929 & 42.4 \\ \hline
No variant & donor-first & 274 & 687 & none \\
           & nondonor-first & 440 & 701 & none\\ \hline
Baseline & donor-first & 468 & 1105 & 42.4 \\
         & nondonor-first & 450 & 808 & 99.5 \\ \hline
Global variant & donor-first & 517 & 1183 & 22.5 \\
         & nondonor-first & 583 & 1368 & 21.6 \\ \hline
Global variant, & donor-first & 467 & 1103 & 42.8 \\
infection days until & nondonor-first & 602 & 1121 & 42.7 \\ 
variant $n = 90,000$  & & & & \\ \hline
Large nondonor & donor-first & 478 & 1151 & 38.9 \\
initial cases,  & donor 75\% & 495 & 1148 & 41.0 \\
 infection days until & donor 50\% & 531 & 1140 & 44.2 \\
 variant $n = 60,000$              & nondonor-first & 498 & 978 & 72.6 \\ \hline
\end{tabular}
\end{table}

\begin{figure}[!ht]
    \centering
    \includegraphics[width=\textwidth]{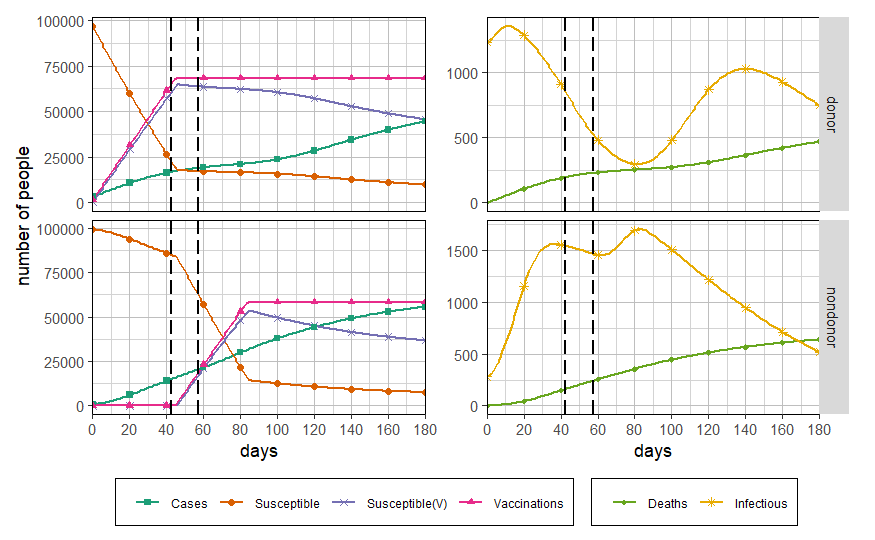}
    \caption{Baseline scenario, donor-first policy}
    \label{fig:time_D}
\end{figure}

\begin{figure}[!ht]
    \centering
    \includegraphics[width=\textwidth]{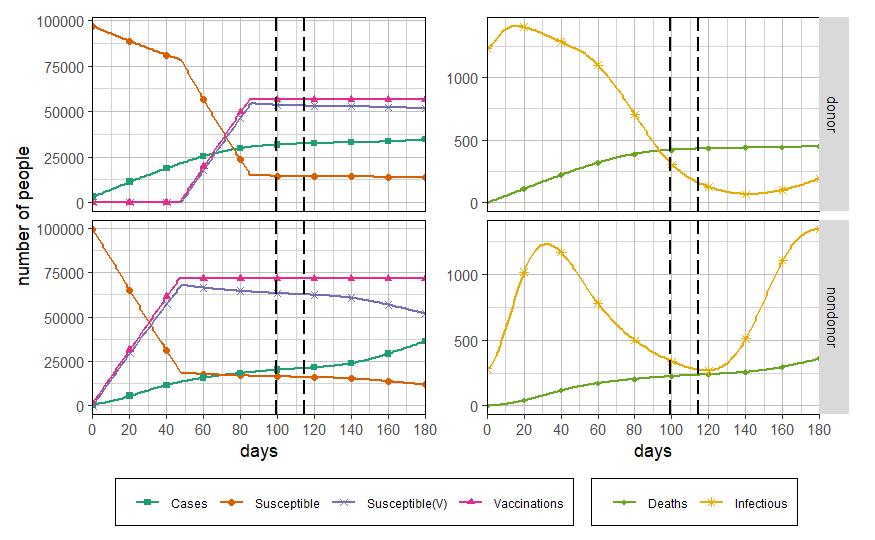}
    \caption{Baseline scenario, nondonor-first policy}
    \label{fig:time_ND}
\end{figure}

The story gets more interesting, however, when we compare vaccine allocation policies in the baseline scenario.  Table \ref{tab:deaths} shows that the nondonor-first policy reduces donor deaths 3.7\%, total deaths 27\%, and delays the variant 57 days compared to the donor-first policy. Results for the baseline scenario under the donor-first and nondonor-first policy are plotted in Figures \ref{fig:time_D} and \ref{fig:time_ND}. Cumulative vaccinations, cases, and deaths, as well as the current susceptible and infectious ($I(t)+I^v(t)$) populations are shown by area. The two vertical lines indicate when the variant arrives in the nondonor and then the donor area. 
For the donor-first policy, the donor area reaches its vaccination limit of 78\% at day 45. After the variant reaches the donor area (day 57), infections rebound, forming a second wave. The nondonor-first policy has fewer donor deaths because, even without vaccination, the first wave is moderated by behavior (fewer contacts) and the variant arrives much later, after both areas are vaccinated. Comparing the nondonor area in Figures \ref{fig:time_D} and \ref{fig:time_ND}, the first surge is much smaller and the second surge later under the nondonor-first policy, resulting in the variant arriving much later. 

The waves can be understood from the herd immunity critical proportions derived by Holleran \textit{et al} \cite{HolleranEtAl_Opt} and restated in Table \ref{tab:herd}. Before the variant, herd immunity in a fully unvaccinated donor area would require a critical (i.e., protected) proportion of 51\%, whereas a fully vaccinated donor area would require a critical proportion of 0\%.  Thus, the actual critical proportion prior to variant emergence is within this range. 
This is reached before roughly day 20, when cases are 12\% and susceptible vaccinated are 30\%. However, by day 80 the variant is dominant in the donor area and the 22\% case rate has not reached herd immunity, even if everyone was vaccinated (33\% is required).

\begin{table}[!h] 
\caption{Herd immunity critical proportions (Holleran \textit{et al.} \cite{HolleranEtAl_Opt})}
\label{tab:herd}
\center
\begin{tabular}{| l | c c | c c | cc|}
\hline 
& \multicolumn{6}{c|}{Critical proportion} \\ \hline
 & \multicolumn{2}{c|}{Before variant} & \multicolumn{2}{c|}{50\% variant} & \multicolumn{2}{c|}{100\% variant} \\
 & Unvacc. & Vacc. & Unvacc. & Vacc. & Unvacc. & Vacc. \\ \hline
 donor area    & 0.51 & 0 & 0.68 & 0.10 & 0.76 & 0.33 \\ 
 nondonor area & 0.57 & 0 & 0.72 & 0.21 & 0.79 & 0.41 \\ \hline
\end{tabular}
\end{table}

Also given in Table \ref{tab:deaths} are the results from the other scenarios. The \say{global variant} scenario counts infections in both areas to determine variant emergence.  In this scenario, the nondonor-first policy does not delay the variant; this is also true when $n$ is doubled to 90,000. For the \say{large nondonor initial cases} scenario, we tested intermediate policies that give 50 or 75\% of vaccinations to the donor. Interestingly, the 50\% policy is \textit{worse} than the nondonor-first policy for donor deaths. Giving 50\% to the nondonor area has little effect on infections in the early days and only delays the variant five days compared to not vaccinating.

\subsection{Sensitivity Analysis}
\label{sec:sensitivity}

This section examines the sensitivity of the best policy and predicted deaths to the more important and uncertain model parameters. Starting with the baseline scenario, parameters were varied one at a time, or in pairs of related parameters. Parameter ranges are listed in Tables \ref{tab:param1}-\ref{tab:param2} and discussed in \ref{sec:param}. 
We first observe in Table \ref{tab:sens}  that the donor-first policy is better for 12 of the 19 parameter changes. The sensitivity of the policy to parameter changes is not surprising, given that the policies only differ by 18 deaths in the baseline scenario. The donor-first policy is also better than nondonor-first in most other scenarios we tested (not reported here), where multiple parameters were changed. 

In several instances, the nondonor-first policy is better because the health risk in the donor country in the first wave is smaller, or the risk in the second wave is larger (low initial infection rate, large change in infection rate for variant, reduced contacts through a smaller $v^u$). In other instances, the nondonor-first policy appears to be better because both areas can be vaccinated more quickly, before the second surge (small proportion willing to be vaccinated, large vaccination budget). 

When the proportion willing to be vaccinated is 100\%, the donor-first policy is better because the second wave is much smaller: as derived by Holleran \textit{et al.} \cite{HolleranEtAl_Opt}, herd immunity in the donor country is reached when everyone is vaccinated and cases reach 33\%; see Table \ref{tab:herd}. In several instances where the donor-first policy is better, the health risk to the donor country in the first wave is larger or the risk in the second wave is smaller (high initial infection rate, longer time in the infectious state, smaller infection rate for variant, increased contacts through a larger $v^u$, reduced vaccine effectiveness through larger $p^e$ and $p^r$). When vaccination takes longer (100\% willing to be vaccinated, smaller vaccination budget), the donor-first policy is also better.

When the nondonor infection rate is doubled ($\gamma = 2$), the critical proportion in the nondonor area, vaccinated and before the variant, changes from 0\% to 41\%. Under a nondonor-first policy, the variant emerges much earlier than in the baseline, making the donor-first policy preferable.

Interestingly, the policy is non-monotone in $r_0$ and vaccine effectiveness. Reducing the time in the infectious state prevents the variant from appearing in the 180 day timeframe, even for the donor-first policy, so it is preferred. Increasing vaccine effectiveness (smaller $p^e$, $p^r$) increases the benefit of the donor-first policy in the first wave enough that it is preferred.

Lastly, we note that the nondonor-first policy in the baseline scenario loses its advantage as the time horizon, $T$, is extended.  The nondonor-first policy postpones but does not prevent second-wave deaths.  Nonetheless, reducing near-term deaths and cases is valuable during a pandemic: it avoids overwhelming hospitals and allows more time for the development of treatments that reduce case mortality.



\begin{table}[!h] 
\caption{Sensitivity analysis for policies and deaths}
\label{tab:sens}
\center
\begin{tabular}{| l | l | l | cc | c |}
\hline 
& & Best & \multicolumn{2}{c|}{Deaths} & Variant \\
Parameter (baseline) & Value & Policy & Donor & Total & (days) \\ 
    \hline \hline
Baseline scenario & & nondonor & 450 & 808 & 99.5  \\ \hline
Proportion willing to vacc. $\rho_a$ (0.78) 
    & 0.5 & nondonor & 659 & 1309 & 63.6 \\ 
    & 1 & donor & 294 & 888 & 42.4 \\ \hline 
Donor initial infection rate $\rho^I$ (0.0036) 
    & 0.0026 & nondonor & 432 & 682 & 153.2 \\
Nondonor is kept at 20\% of donor 
    & 0.0049 & donor & 480 & 1127 & 39.8 \\ \hline
Time in infectious state $1/r_0$ (3.9 days) 
    & 2.5 & donor & 173 & 430 & none \\
    & 5.1 & donor & 569 & 1250 & 33.93 \\  \hline
Testing effect $\Delta r_a$, donor area (0.035) 
    & 0.026 & nondonor & 459 & 816 & 99.5 \\
    & 0.060 & donor & 405 & 1043 & 42.4 \\ \hline
Infection multiplier, nondonor $\gamma_a$ (1) 
    & 2 & donor & 506 & 1434 & 27.6 \\ \hline
Change in infection rate $\Delta a$ (0.6) 
    & 0.3 & donor & 343 & 875 & 42.4 \\
for variant & 0.9 & nondonor & 464 & 881 & 99.5 \\ \hline
Behavior dynamics $v^u$ (0.03) 
    & 0.025 & nondonor & 395 & 632 & 178.2 \\
    & 0.10 & donor & 621 & 1521 & 32.1 \\ \hline
Vaccine effectiveness $p^e, p^r$ (both 0.6) 
    & 0.5 & donor & 349 & 928 & 42.4 \\
    & 0.8 & donor & 621 & 1339 & 42.4 \\ \hline
Vaccination budget $B$ (1500) 
    & 1000 & donor & 502 & 1223 & 42.4 \\
per day & 2000 & nondonor & 385 & 621 & 180 \\ \hline
Time horizon $T$ (180) & 270 & nondonor & 561 & 1119  & 99.5 \\
& 360 & donor & 589 & 1292 & 42.4 \\ \hline
\end{tabular}
\end{table}


\section{Conclusions}
\label{sec:conc}

We have demonstrated how the SEIR framework developed by Holleran \textit{et al.} \cite{HolleranEtAl_Opt} can provide qualitative insights into conditions under which a high-income country might realize health benefits by donating vaccine doses to a low-income country before its population is fully vaccinated.  

Using the COVID-19 pandemic as a case study, we find that there exist realistic scenarios under which a nondonor-first (altrustic) vaccination policy reduces not only total deaths but deaths within the donor country.  Moreover, a nondonor-first vaccination policy can dramatically postpone the emergence of a variant that could be more contagious and transmissible than the original virus, giving both the donor country and the nondonor country more time to vaccinate their populations.  This is particularly the case under an assumption of self-regulating behavior dynamics wherein a population takes actions (such as social distancing or masking) to reduce transmission during periods of high infections.  Notably, we found a realistic scenario in which donating all vaccines is  better for the donor country than a sharing policy in which half of the vaccines are donated and half are retained. Thus, we have demonstrated that in addition to moral arguments against vaccine nationalism, there exist local public health arguments as well.  


The choice of donor-first or nondonor-first is sensitive to population behavioral characteristics, such as willingness to vaccinate and behavioral responses to local infection rates, and to characteristics of the variant.  The nondonor-first policy outperforms the donor-first policy when the proportion willing to vaccinate is low (equivalently, the vaccination budget is high), the initial infection rate in the donor country is low, and the donor country adopts 
behavioral changes such as social distancing and masking at a low threshold of proportion infectious.  In these instances, we interpret that the donor country should 
donate vaccines when the local health impact of the vaccine is limited.  Additionally, nondonor-first outperforms donor-first when the new variant will be highly contagious, suggesting that the donor country should also 
donate vaccines when delaying emergence of a variant is especially valuable. 

We conclude by noting that some have claimed that COVID-19 vaccination campaigns in low-income regions, such as Africa, may not be in those regions' best interests either. Writes John Johnson, vaccination adviser for Doctors Without Borders, ``Is this the most important thing to try to carry out in countries where there are much bigger problems with malaria, with polio, with measles, with cholera, with meningitis, with malnutrition? Is this what we want to spend our resources on in those countries?  Because at this point, it’s not for those people: It’s to try to prevent new variants,'' \cite{NYT_2022_CovidAfricaDeaths}.  Thus, while vaccine inequity perpetuates disparate health outcomes globally, an emphasis on vaccine donation can itself be fraught with donor self-interest.  Our model permits a 
comparison of multiple policies to minimize pandemic-related deaths in the donor country and in total so that such questions of equity can be thoroughly examined.

\bibliographystyle{elsarticle-num-names} 
\bibliography{references}

\begin{thebibliography}{48}
\expandafter\ifx\csname natexlab\endcsname\relax\def\natexlab#1{#1}\fi
\providecommand{\url}[1]{\texttt{#1}}
\providecommand{\href}[2]{#2}
\providecommand{\path}[1]{#1}
\providecommand{\DOIprefix}{doi:}
\providecommand{\ArXivprefix}{arXiv:}
\providecommand{\URLprefix}{URL: }
\providecommand{\Pubmedprefix}{pmid:}
\providecommand{\doi}[1]{\href{http://dx.doi.org/#1}{\path{#1}}}
\providecommand{\Pubmed}[1]{\href{pmid:#1}{\path{#1}}}
\providecommand{\bibinfo}[2]{#2}
\ifx\xfnm\relax \def\xfnm[#1]{\unskip,\space#1}\fi
\bibitem[{Clinton and Yoo(2022)}]{ClintonYoo}
\bibinfo{author}{C.~Clinton}, \bibinfo{author}{K.~J. Yoo},
\newblock \bibinfo{title}{Is {COVAX} to blame for failing to close global
  vaccination disparities?},
\newblock \bibinfo{journal}{Health Affairs Forefront}  (\bibinfo{year}{June 14,
  2022}). \bibinfo{note}{Accessed at
  \url{https://www.healthaffairs.org/do/10.1377/forefront.20220609.695589} on
  10 January 2023.}
\bibitem[{Holleran et~al.(2023)Holleran, Martonosi, and
  Veatch}]{HolleranEtAl_Opt}
\bibinfo{author}{A.~Holleran}, \bibinfo{author}{S.~E. Martonosi},
  \bibinfo{author}{M.~Veatch},
\newblock \bibinfo{title}{International vaccine allocation: An optimization
  framework}  (\bibinfo{year}{2023}).
  \bibinfo{note}{\url{https://arxiv.org/abs/2303.05917}. Preprint.}
\bibitem[{Thomas(2020)}]{NYT_PF95}
\bibinfo{author}{K.~Thomas}, \bibinfo{title}{New {P}fizer results: Coronavirus
  vaccine is safe and 95\% effective},
  \bibinfo{howpublished}{\url{https://www.nytimes.com/2020/11/18/health/pfizer-covid-vaccine.html}},
  \bibinfo{year}{2020}. \bibinfo{note}{Accessed 8 Dec. 2020}.
\bibitem[{Grady(2020)}]{moderna}
\bibinfo{author}{D.~Grady}, \bibinfo{title}{Early data show {M}oderna’s
  {C}oronavirus vaccine is 94.5\% effective}, \bibinfo{howpublished}{New York
  Times.
  \url{https://www.nytimes.com/2020/11/16/health/Covid-moderna-vaccine.html}},
  \bibinfo{year}{2020}. \bibinfo{note}{Accessed 16 Nov. 2020}.
\bibitem[{{Businesswire}(2020{\natexlab{a}})}]{pa}
\bibinfo{author}{{Businesswire}}, \bibinfo{title}{Pfizer and {BioNTech}
  announce an agreement with {U.S.} government for up to 600 million doses of
  {mRNA}-based vaccine candidate against {SARS-CoV-2}},
  \bibinfo{howpublished}{\url{https://www.pfizer.com/news/press-release/press-release-detail/pfizer-and-biontech-announce-agreement-us-government-600}},
  \bibinfo{year}{2020}{\natexlab{a}}. \bibinfo{note}{Accessed 1 Dec. 2020}.
\bibitem[{{Businesswire}(2020{\natexlab{b}})}]{ma}
\bibinfo{author}{{Businesswire}}, \bibinfo{title}{Moderna announces supply
  agreement with {U.S.} government for initial 100 million doses of {mRNA}
  vaccine against {COVID-19} ({mRNA}-1273)},
  \bibinfo{howpublished}{\url{https://www.businesswire.com/news/home/20200811005852/en/}},
  \bibinfo{year}{2020}{\natexlab{b}}. \bibinfo{note}{Accessed 1 Dec. 2020}.
\bibitem[{{GAVI}(2022)}]{Gavi}
\bibinfo{author}{{GAVI}}, \bibinfo{title}{{COVAX} dose donation table},
  \bibinfo{howpublished}{\url{https://www.gavi.org/news/document-library/covax-dose-donation-table}},
  \bibinfo{year}{2022}. \bibinfo{note}{Accessed 18 Aug. 2022}.
\bibitem[{{World Health Organization}(2022)}]{WHO}
\bibinfo{author}{{World Health Organization}}, \bibinfo{title}{Vaccine equity},
  \bibinfo{year}{2022}. \bibinfo{note}{Accessed at
  \url{https://www.who.int/campaigns/vaccine-equity} on 7 June 2022.}
\bibitem[{Hunter et~al.(2022)Hunter, Karim, Baden, Farrar, Hamel, Longo
  et~al.}]{HunterEtAl_2022_NEJM}
\bibinfo{author}{D.~J. Hunter}, \bibinfo{author}{S.~S.~A. Karim},
  \bibinfo{author}{L.~R. Baden}, \bibinfo{author}{J.~J. Farrar},
  \bibinfo{author}{M.~B. Hamel}, \bibinfo{author}{D.~L. Longo}, et~al.,
\newblock \bibinfo{title}{Addressing vaccine inequity — {Covid-19} vaccines
  as a global public good},
\newblock \bibinfo{journal}{New England Journal of Medicine}
  \bibinfo{volume}{386} (\bibinfo{year}{2022}) \bibinfo{pages}{1176--1179}.
  \bibinfo{note}{\url{https://www.nejm.org/doi/full/10.1056/NEJMe2202547}}.
\bibitem[{Asundi et~al.(2021)Asundi, O'Leary, and
  Bhadelia}]{AsundiOLearyBhadelia_2021_CHM}
\bibinfo{author}{A.~Asundi}, \bibinfo{author}{C.~O'Leary},
  \bibinfo{author}{N.~Bhadelia},
\newblock \bibinfo{title}{Global {COVID-19} vaccine inequity: The scope, the
  impact, and the challenges},
\newblock \bibinfo{journal}{Cell Host Microbe} \bibinfo{volume}{29}
  (\bibinfo{year}{2021}) \bibinfo{pages}{1036--1039}.
\bibitem[{Ghebreyesus(2021)}]{Ghebreyesus_2021_PLOS}
\bibinfo{author}{T.~A. Ghebreyesus},
\newblock \bibinfo{title}{Five steps to solving the vaccine inequity crisis},
\newblock \bibinfo{journal}{PLOS Global Public Health} \bibinfo{volume}{1}
  (\bibinfo{year}{2021}).
  \bibinfo{note}{\url{https://doi.org/10.1371/journal.pgph.0000032}}.
\bibitem[{Hinshaw(2021)}]{WSJ}
\bibinfo{author}{D.~Hinshaw},
\newblock \bibinfo{title}{Omicron variant highlights risks of low vaccination
  rates in poor countries},
\newblock \bibinfo{journal}{Wall Street Journal}  (\bibinfo{year}{2021}).
  \bibinfo{note}{Accessed at
  \url{https://www.wsj.com/articles/omicron-variant-highlights-risks-of-low-vaccination-rates-in-poor-countries-11637960558}
  on 7 June 2022.}
\bibitem[{Houtman et~al.(2021)Houtman, Shultz, Rivera, Bass, Bright, Luo, and
  Gilmour}]{Rockefeller}
\bibinfo{author}{J.~Houtman}, \bibinfo{author}{L.~Shultz},
  \bibinfo{author}{J.~M. Rivera}, \bibinfo{author}{E.~Bass},
  \bibinfo{author}{R.~A. Bright}, \bibinfo{author}{D.~Luo},
  \bibinfo{author}{J.~Gilmour}, \bibinfo{title}{Vaccine inequity increases the
  risk of new {SARS-CoV-2} variants emerging},
  \bibinfo{howpublished}{\url{https://www.rockefellerfoundation.org/case-study/vaccine-inequity-increases-the-risk-of-new-sars-cov-2-variants-emerging/}},
  \bibinfo{year}{2021}. \bibinfo{note}{Accessed 18 Aug. 2022}.
\bibitem[{Volpert et~al.(2021)Volpert, Banerjee, and
  Sharma}]{VolpertVitalySharma_EcoComp_2021}
\bibinfo{author}{V.~Volpert}, \bibinfo{author}{M.~Banerjee},
  \bibinfo{author}{S.~Sharma},
\newblock \bibinfo{title}{Epidemic progression and vaccination in a
  heterogeneous population. application to the {C}ovid-19 epidemic},
\newblock \bibinfo{journal}{Ecological Complexity} \bibinfo{volume}{47}
  (\bibinfo{year}{2021}). \DOIprefix\doi{10.1016/j.ecocom.2021.100940}.
\bibitem[{Dolbeault and Turinici(2020)}]{DolbeaultTurinici_MathMod_2020}
\bibinfo{author}{J.~Dolbeault}, \bibinfo{author}{G.~Turinici},
\newblock \bibinfo{title}{Heterogeneous social interactions and the {COVID-19}
  lockdown outcome in a multi-group {SEIR} model},
\newblock \bibinfo{journal}{Mathematical Modelling of Natural Phenomena}
  \bibinfo{volume}{15} (\bibinfo{year}{2020}).
  \DOIprefix\doi{10.1051/mmnp/2020025}.
\bibitem[{Bicher et~al.(2022)Bicher, Rippinger, Zechmeister, Jahn, Sroczynski,
  Muehlberger et~al.}]{BicherEtAl_PLOS_2022}
\bibinfo{author}{M.~Bicher}, \bibinfo{author}{C.~Rippinger},
  \bibinfo{author}{M.~Zechmeister}, \bibinfo{author}{B.~Jahn},
  \bibinfo{author}{G.~Sroczynski}, \bibinfo{author}{N.~Muehlberger}, et~al.,
\newblock \bibinfo{title}{An iterative algorithm for optimizing {COVID-19}
  vaccination strategies considering unknown supply},
\newblock \bibinfo{journal}{PLOS One} \bibinfo{volume}{17}
  (\bibinfo{year}{2022}). \DOIprefix\doi{10.1371/journal.pone.0265957}.
\bibitem[{Pan et~al.(2022)Pan, Ng, Dong, and Cheng}]{PanEtAl_AOR_2022}
\bibinfo{author}{Y.~Pan}, \bibinfo{author}{C.~T. Ng},
  \bibinfo{author}{C.~Dong}, \bibinfo{author}{T.~C.~E. Cheng},
\newblock \bibinfo{title}{Information sharing and coordination in a vaccine
  supply chain},
\newblock \bibinfo{journal}{Annals of Operations Research}  (\bibinfo{year}{FEB
  2022}). \DOIprefix\doi{10.1007/s10479-022-04562-1}.
\bibitem[{Tavana et~al.(2021)Tavana, Govindan, Nasr, Heidary, and
  Mina}]{TavanaEtAl_AOR_2021}
\bibinfo{author}{M.~Tavana}, \bibinfo{author}{K.~Govindan},
  \bibinfo{author}{A.~K. Nasr}, \bibinfo{author}{M.~S. Heidary},
  \bibinfo{author}{H.~Mina},
\newblock \bibinfo{title}{A mathematical programming approach for equitable
  {COVID-19} vaccine distribution in developing countries},
\newblock \bibinfo{journal}{Annals of Operations Research}  (\bibinfo{year}{JUN
  2021}). \DOIprefix\doi{10.1007/s10479-021-04130-z}.
\bibitem[{Van~Oorschot et~al.(2022)Van~Oorschot, Van~Wassenhove, and
  Jahre}]{VanOorschotEtAl_2022}
\bibinfo{author}{K.~E. Van~Oorschot}, \bibinfo{author}{L.~N. Van~Wassenhove},
  \bibinfo{author}{M.~Jahre},
\newblock \bibinfo{title}{Collaboration–competition dilemma in flattening the
  {COVID-19} curve},
\newblock \bibinfo{journal}{Production \& Operations Management}
  (\bibinfo{year}{2022}) \bibinfo{pages}{1--17}.
  \DOIprefix\doi{https://doi.org/10.1111/poms.13709}.
\bibitem[{Duijzer et~al.(2018)Duijzer, {van Jaarsveld}, Wallinga, and
  Dekker}]{DuijzerEtAl_2018_POM}
\bibinfo{author}{L.~E. Duijzer}, \bibinfo{author}{W.~L. {van Jaarsveld}},
  \bibinfo{author}{J.~Wallinga}, \bibinfo{author}{R.~Dekker},
\newblock \bibinfo{title}{Dose-optimal vaccine allocation over multiple
  populations},
\newblock \bibinfo{journal}{Production \& Operations Management}
  \bibinfo{volume}{27} (\bibinfo{year}{2018}) \bibinfo{pages}{143--159}.
\bibitem[{Rotesi et~al.(2021)Rotesi, Pin, Cucciniello, Malik, Paintsil,
  Bokemper et~al.}]{RotesiEtAl2021}
\bibinfo{author}{T.~Rotesi}, \bibinfo{author}{P.~Pin},
  \bibinfo{author}{M.~Cucciniello}, \bibinfo{author}{A.~A. Malik},
  \bibinfo{author}{E.~E. Paintsil}, \bibinfo{author}{S.~E. Bokemper}, et~al.,
\newblock \bibinfo{title}{National interest may require distributing
  {COVID‑19} vaccines to other countries},
\newblock \bibinfo{journal}{Scientific Reports} \bibinfo{volume}{11}
  (\bibinfo{year}{2021}) \bibinfo{pages}{18253}.
  \bibinfo{note}{\url{https://www.nature.com/articles/s41598-021-97544-5}}.
\bibitem[{Nolen(2022)}]{NYT_2022_CovidAfricaDeaths}
\bibinfo{author}{S.~Nolen}, \bibinfo{title}{Trying to solve a {COVID} mystery:
  {A}frica’s low death rates}, \bibinfo{howpublished}{New York Times
  \url{https://www.nytimes.com/2022/03/23/health/covid-africa-deaths.html}},
  \bibinfo{year}{2022}. \bibinfo{note}{Accessed on 19 Aug. 2022.}
\bibitem[{Rubio-Herrero and Wang(2022)}]{Rubio-HerreroWang_CIE_2022}
\bibinfo{author}{J.~Rubio-Herrero}, \bibinfo{author}{Y.~Wang},
\newblock \bibinfo{title}{A flexible rolling regression framework for the
  identification of time-varying {SIRD} models},
\newblock \bibinfo{journal}{Computers \& Industrial Engineering}
  \bibinfo{volume}{167} (\bibinfo{year}{2022}).
  \DOIprefix\doi{10.1016/j.cie.2022.108003}.
\bibitem[{Gallo et~al.(2022)Gallo, Frasca, Latora, and
  Russo}]{GalloEtAl_ScienceAdv_2022}
\bibinfo{author}{L.~Gallo}, \bibinfo{author}{M.~Frasca},
  \bibinfo{author}{V.~Latora}, \bibinfo{author}{G.~Russo},
\newblock \bibinfo{title}{Lack of practical identifiability may hamper reliable
  predictions in {COVID-19} epidemic models},
\newblock \bibinfo{journal}{Science Advances} \bibinfo{volume}{8}
  (\bibinfo{year}{2022}). \DOIprefix\doi{10.1126/sciadv.abg5234}.
\bibitem[{Koenen et~al.(2021)Koenen, Balvert, Brekelmans, Fleuren, Stienen, and
  Wagenaar}]{KoenenEtAl_PLOS_2021}
\bibinfo{author}{M.~Koenen}, \bibinfo{author}{M.~Balvert},
  \bibinfo{author}{R.~Brekelmans}, \bibinfo{author}{H.~Fleuren},
  \bibinfo{author}{V.~Stienen}, \bibinfo{author}{J.~Wagenaar},
\newblock \bibinfo{title}{Forecasting the spread of sars-cov-2 is inherently
  ambiguous given the current state of virus research},
\newblock \bibinfo{journal}{PLOS One} \bibinfo{volume}{16}
  (\bibinfo{year}{2021}). \DOIprefix\doi{10.1371/journal.pone.0245519}.
\bibitem[{Nikolopoulos et~al.(2021{\natexlab{a}})Nikolopoulos, Tsinopoulos, and
  Vasilakis}]{NikolopoulosEtAl_JORS_2021}
\bibinfo{author}{K.~Nikolopoulos}, \bibinfo{author}{C.~Tsinopoulos},
  \bibinfo{author}{C.~Vasilakis},
\newblock \bibinfo{title}{Operational research in the time of {COVID-19}: The
  `science for better' or worse in the absence of hard data},
\newblock \bibinfo{journal}{Journal of the Operational Research Society}
  (\bibinfo{year}{MAY 2021}{\natexlab{a}}).
  \DOIprefix\doi{10.1080/01605682.2021.1930208}.
\bibitem[{Nikolopoulos et~al.(2021{\natexlab{b}})Nikolopoulos, Punia, Schafers,
  Tsinopoulos, and Vasilakis}]{NikolopoulosEtAl_EJOR_2021}
\bibinfo{author}{K.~Nikolopoulos}, \bibinfo{author}{S.~Punia},
  \bibinfo{author}{A.~Schafers}, \bibinfo{author}{C.~Tsinopoulos},
  \bibinfo{author}{C.~Vasilakis},
\newblock \bibinfo{title}{Forecasting and planning during a pandemic:
  {COVID-19} growth rates, supply chain disruptions, and governmental
  decisions},
\newblock \bibinfo{journal}{European Journal of Operational Research}
  \bibinfo{volume}{290} (\bibinfo{year}{2021}{\natexlab{b}})
  \bibinfo{pages}{99--115}. \DOIprefix\doi{10.1016/j.ejor.2020.08.001}.
\bibitem[{Ndugga et~al.(2022)Ndugga, Hill, Artiga, and Haldar}]{KFF}
\bibinfo{author}{N.~Ndugga}, \bibinfo{author}{L.~Hill},
  \bibinfo{author}{S.~Artiga}, \bibinfo{author}{S.~Haldar},
  \bibinfo{title}{Latest data on {COVID-19} vaccinations by race/ethnicity},
  \bibinfo{howpublished}{Kaiser Family Foundation
  \url{https://www.kff.org/coronavirus-covid-19/issue-brief/latest-data-on-covid-19-vaccinations-by-race-ethnicity/}},
  \bibinfo{year}{2022}. \bibinfo{note}{Accessed on 25 Aug. 2022.}
\bibitem[{Arce et~al.(2021)Arce, Warren, Meriggi, Scacco, McMurry, Voors
  et~al.}]{Nature}
\bibinfo{author}{J.~S.~S. Arce}, \bibinfo{author}{S.~S. Warren},
  \bibinfo{author}{N.~F. Meriggi}, \bibinfo{author}{A.~Scacco},
  \bibinfo{author}{N.~McMurry}, \bibinfo{author}{M.~Voors}, et~al.,
\newblock \bibinfo{title}{{COVID-19} vaccine acceptance and hesitancy in low-
  and middle-income countries},
\newblock \bibinfo{journal}{Nature Medicine} \bibinfo{volume}{27}
  (\bibinfo{year}{2021}) \bibinfo{pages}{1385--1394}. \bibinfo{note}{Accessed
  on 30 August 2022 \url{https://www.nature.com/articles/s41591-021-01454-y}}.
\bibitem[{Nicholson et~al.(2022)Nicholson, Lehmann, Padellini, Pouwels,
  Jersakova, Lomax et~al.}]{Nature_Eng_prev}
\bibinfo{author}{G.~Nicholson}, \bibinfo{author}{B.~Lehmann},
  \bibinfo{author}{T.~Padellini}, \bibinfo{author}{K.~B. Pouwels},
  \bibinfo{author}{R.~Jersakova}, \bibinfo{author}{J.~Lomax}, et~al.,
\newblock \bibinfo{title}{Improving local prevalence estimates of {SARS-CoV-2}
  infections using a causal debiasing framework},
\newblock \bibinfo{journal}{Nature Microbiology} \bibinfo{volume}{7}
  (\bibinfo{year}{2022}) \bibinfo{pages}{97--107}. \bibinfo{note}{Accessed on
  30 Aug. 2022. \url{https://doi.org/10.1038/s41564-021-01029-0}}.
\bibitem[{Chiu and Ndeffo-Mbah(2021)}]{PLOS_US_prev}
\bibinfo{author}{W.~A. Chiu}, \bibinfo{author}{M.~L. Ndeffo-Mbah},
\newblock \bibinfo{title}{Using test positivity and reported case rates to
  estimate state-level {COVID-19} prevalence and seroprevalence in the {U}nited
  {S}tates},
\newblock \bibinfo{journal}{PLOS Computational Biology} \bibinfo{volume}{17}
  (\bibinfo{year}{2021}) \bibinfo{pages}{1--19}. \URLprefix
  \url{https://doi.org/10.1371/journal.pcbi.1009374}.
\bibitem[{Kalish et~al.(2021)Kalish, Klumpp-Thomas, Hunsberger, Baus, Fay,
  Siripong et~al.}]{Science_US_prev}
\bibinfo{author}{H.~Kalish}, \bibinfo{author}{C.~Klumpp-Thomas},
  \bibinfo{author}{S.~Hunsberger}, \bibinfo{author}{H.~A. Baus},
  \bibinfo{author}{M.~P. Fay}, \bibinfo{author}{N.~Siripong}, et~al.,
\newblock \bibinfo{title}{Undiagnosed {SARS-CoV-2} seropositivity during the
  first 6 months of the {COVID-19} pandemic in the {U}nited {S}tates},
\newblock \bibinfo{journal}{Science Translational Medicine}
  \bibinfo{volume}{13} (\bibinfo{year}{2021}) \bibinfo{pages}{eabh3826}.
  \DOIprefix\doi{10.1126/scitranslmed.abh3826}.
\bibitem[{Hamadeh et~al.(2021)Hamadeh, Feng, J, and Wong}]{JMIR_Canada_prev}
\bibinfo{author}{A.~Hamadeh}, \bibinfo{author}{Z.~Feng}, \bibinfo{author}{J.~N.
  J}, \bibinfo{author}{W.~W. Wong},
\newblock \bibinfo{title}{Estimation of {COVID-19} period prevalence and the
  undiagnosed population in {C}anadian provinces: Model-based analysis},
\newblock \bibinfo{journal}{JMIR Public Health Surveillance}
  \bibinfo{volume}{7} (\bibinfo{year}{2021}) \bibinfo{pages}{e26409}.
  \bibinfo{note}{Accessed on 30 Aug. 2022.
  \url{https://publichealth.jmir.org/2021/9/e26409}}.
\bibitem[{Byrne et~al.(2020)Byrne, McEvoy, Collins, Hunt, Casey, Barber
  et~al.}]{Byrne}
\bibinfo{author}{A.~W. Byrne}, \bibinfo{author}{D.~McEvoy},
  \bibinfo{author}{A.~B. Collins}, \bibinfo{author}{K.~Hunt},
  \bibinfo{author}{M.~Casey}, \bibinfo{author}{A.~Barber}, et~al.,
\newblock \bibinfo{title}{Inferred duration of infectious period of
  \text{SARS-CoV-2}: rapid scoping review and analysis of available evidence
  for asymptomatic and symptomatic \text{COVID}-19 cases} \bibinfo{volume}{10}
  (\bibinfo{year}{2020}). \DOIprefix\doi{10.1136/bmjopen-2020-039856}.
\bibitem[{Aronna et~al.(2021)Aronna, Guglielmi, and
  Moschen}]{AronnaEtAl_2021_Epidemics}
\bibinfo{author}{M.~Aronna}, \bibinfo{author}{R.~Guglielmi},
  \bibinfo{author}{L.~Moschen},
\newblock \bibinfo{title}{A model for {COVID-19} with isolation, quarantine and
  testing as control measures},
\newblock \bibinfo{journal}{Epidemics} \bibinfo{volume}{34}
  (\bibinfo{year}{2021}) \bibinfo{pages}{100437}. \bibinfo{note}{Accessed on 9
  September 2022
  \url{https://www.sciencedirect.com/science/article/pii/S1755436521000025}}.
\bibitem[{Rasmussen and Popescu(2021)}]{RasmussenPopescu}
\bibinfo{author}{A.~L. Rasmussen}, \bibinfo{author}{S.~V. Popescu},
\newblock \bibinfo{title}{{SARS-CoV-2} transmission without symptoms:
  Symptomless transmission silently drives viral spread and is key to ending
  the pandemic},
\newblock \bibinfo{journal}{Science} \bibinfo{volume}{371}
  (\bibinfo{year}{2021}) \bibinfo{pages}{1206--1207}. \bibinfo{note}{Accessed
  on 26 September 2022
  \url{https://www.science.org/doi/epdf/10.1126/science.abf9569}}.
\bibitem[{{Centers for Disease Control and
  Prevention}(2022)}]{CDC_durationisolation}
\bibinfo{author}{{Centers for Disease Control and Prevention}},
  \bibinfo{title}{Ending isolation and precautions for people with {COVID-19}:
  Interim guidance}, \bibinfo{year}{2022}. \bibinfo{note}{Accessed on 30
  September 2022
  \url{https://www.cdc.gov/coronavirus/2019-ncov/hcp/duration-isolation.html}}.
\bibitem[{Lauer et~al.(2020)Lauer, Grantz, Bi, Jones, Zheng, Meredith
  et~al.}]{Lauer}
\bibinfo{author}{S.~A. Lauer}, \bibinfo{author}{K.~H. Grantz},
  \bibinfo{author}{Q.~Bi}, \bibinfo{author}{F.~K. Jones},
  \bibinfo{author}{Q.~Zheng}, \bibinfo{author}{H.~R. Meredith}, et~al.,
\newblock \bibinfo{title}{The incubation period of coronavirus disease 2019
  (\text{COVID}-19) from publicly reported confirmed cases: Estimation and
  application},
\newblock \bibinfo{journal}{Annals of Internal Medicine} \bibinfo{volume}{172}
  (\bibinfo{year}{2020}) \bibinfo{pages}{577--582}. \URLprefix
  \url{https://doi.org/10.7326/M20-0504}. \DOIprefix\doi{10.7326/M20-0504}.
  \href{http://arxiv.org/abs/https://doi.org/10.7326/M20-0504}{{\tt
  arXiv:https://doi.org/10.7326/M20-0504}}, \bibinfo{note}{pMID: 32150748}.
\bibitem[{Johnson et~al.(2022)Johnson, Amin, Ali, Hoots, Cadwell, Arora
  et~al.}]{CDC_mortality}
\bibinfo{author}{A.~G. Johnson}, \bibinfo{author}{A.~B. Amin},
  \bibinfo{author}{A.~R. Ali}, \bibinfo{author}{B.~Hoots},
  \bibinfo{author}{B.~L. Cadwell}, \bibinfo{author}{S.~Arora}, et~al.,
\newblock \bibinfo{title}{{COVID-19} incidence and death rates among
  unvaccinated and fully vaccinated adults with and without booster doses
  during periods of delta and omicron variant emergence — 25 {U.S.}
  jurisdictions, {A}pril 4–{D}ecember 25, 2021},
\newblock \bibinfo{journal}{Morbidity and Mortality Weekly Report}
  \bibinfo{volume}{71} (\bibinfo{year}{2022}) \bibinfo{pages}{132--138}.
  \bibinfo{note}{Accessed on 6 December 2022 at
  \url{https://www.cdc.gov/mmwr/volumes/71/wr/mm7104e2.htm}}.
\bibitem[{Hansen(2021)}]{Hansen}
\bibinfo{author}{P.~R. Hansen}, \bibinfo{title}{Relative contagiousness of
  emerging virus variants: An analysis of the alpha, delta, and omicron
  {SARS-CoV-2} variants}, \bibinfo{year}{2021}. \URLprefix
  \url{https://arxiv.org/abs/2110.00533}.
  \DOIprefix\doi{10.48550/ARXIV.2110.00533}, \bibinfo{note}{preprint.}
\bibitem[{Katella(2022)}]{Yale}
\bibinfo{author}{K.~Katella}, \bibinfo{title}{Omicron, delta, alpha, and more:
  What to know about the coronavirus variants}, \bibinfo{year}{2022}.
  \bibinfo{note}{Accessed on 17 November 2022
  \url{https://www.yalemedicine.org/news/covid-19-variants-of-concern-omicron}}.
\bibitem[{Sun et~al.(2022)Sun, Keating, and Achenbach}]{Wapo}
\bibinfo{author}{L.~H. Sun}, \bibinfo{author}{D.~Keating},
  \bibinfo{author}{J.~Achenbach}, \bibinfo{title}{Coronavirus has infected
  majority of {A}mericans, blood tests indicate}, \bibinfo{year}{2022}.
  \bibinfo{note}{Accessed on 6 December 2022 at
  \url{https://www.washingtonpost.com/health/2022/04/26/majority-americans-coronavirus-infections/}}.
\bibitem[{Eyre et~al.(2022)Eyre, Taylor, Purver, Chapman, Fowler, Pouwels
  et~al.}]{rr}
\bibinfo{author}{D.~W. Eyre}, \bibinfo{author}{D.~Taylor},
  \bibinfo{author}{M.~Purver}, \bibinfo{author}{D.~Chapman},
  \bibinfo{author}{T.~Fowler}, \bibinfo{author}{K.~B. Pouwels}, et~al.,
\newblock \bibinfo{title}{Effect of {COVID}-19 vaccination on transmission of
  alpha and delta variants},
\newblock \bibinfo{journal}{New England Journal of Medicine}
  \bibinfo{volume}{386} (\bibinfo{year}{2022}) \bibinfo{pages}{744--756}.
  \URLprefix \url{https://doi.org/10.1056/NEJMoa2116597}.
  \DOIprefix\doi{10.1056/NEJMoa2116597}.
  \href{http://arxiv.org/abs/https://doi.org/10.1056/NEJMoa2116597}{{\tt
  arXiv:https://doi.org/10.1056/NEJMoa2116597}}.
\bibitem[{{World Health Organization}(2021)}]{whoomicron}
\bibinfo{author}{{World Health Organization}}, \bibinfo{title}{Classification
  of {Omicron (B.1.1.529): SARS-CoV-2} variant of concern},
  \bibinfo{year}{2021}. \bibinfo{note}{Accessed on 21 November 2022
  \url{https://www.who.int/news/item/26-11-2021-classification-of-omicron-(b.1.1.529)-sars-cov-2-variant-of-concern}}.
\bibitem[{{Wikipedia contributors}(2022)}]{wikiomicron}
\bibinfo{author}{{Wikipedia contributors}}, \bibinfo{title}{Timeline of the
  {SARS-CoV-2} omicron variant}, \bibinfo{year}{2022}. \bibinfo{note}{Accessed
  on 21 November 2022
  \url{https://en.wikipedia.org/w/index.php?title=Timeline_of_the_SARS-CoV-2_Omicron_variant&oldid=1122899360}}.
\bibitem[{Craig(2021)}]{msn}
\bibinfo{author}{E.~Craig}, \bibinfo{title}{Omicron becomes dominant in {South
  Africa} in just one week}, \bibinfo{year}{2021}. \bibinfo{note}{Accessed on
  21 November 2022
  \url{https://www.msn.com/en-gb/health/medical/omicron-becomes-dominant-in-south-africa-in-just-one-week/ar-AARoeLC}}.
\bibitem[{{Korea Times}(2022)}]{koreatimes}
\bibinfo{author}{{Korea Times}}, \bibinfo{title}{Omicron becomes dominant
  variant; new {COVID-19} cases above 7,000 for 3rd day}, \bibinfo{year}{2022}.
  \bibinfo{note}{Accessed on 21 November 2022
  \url{https://www.koreatimes.co.kr/www/nation/2022/01/119_322761.html}}.
\bibitem[{{COVID-19 Forecasting Team}(2023)}]{LancetNews}
\bibinfo{author}{{COVID-19 Forecasting Team}},
\newblock \bibinfo{journal}{The Lancet} \bibinfo{volume}{401}
  (\bibinfo{year}{2023}) \bibinfo{pages}{833--842}.
  \DOIprefix\doi{https://doi.org/10.1016/S0140-6736(22)02465-5}.

\end{thebibliography}

\appendix

\section{Parameter Estimation}
\label{sec:param}

Data estimation is a particular challenge in COVID-19 modeling, due to asymptomatic carriers \cite{Rubio-HerreroWang_CIE_2022, GalloEtAl_ScienceAdv_2022}, sensitivity to model assumptions \cite{KoenenEtAl_PLOS_2021}, and the absence of accurate epidemiological data during the pandemic \cite{NikolopoulosEtAl_JORS_2021, NikolopoulosEtAl_EJOR_2021}.   To test and evaluate 
SEIR-V we construct a baseline scenario with two areas, one representing donor countries and the other representing low-income recipient (nondonor) countries. 
We also examine its results on a wide range of parameter settings.  
This section describes our choice of base and sensitivity range values for each parameter. 

\subsection{Parameters that Depend on the Area (Table \ref{tab:param1})} 
\label{ssec:paramGenPop}

Table \ref{tab:param1} lists parameters of SEIR-V that vary by geographic area. 
These parameters are estimated as follows.

\begin{itemize}

\item \textbf{$N_a$ (initial population)}:  For simplicity, 
we assume an equal initial population of 100,000 in both areas.  

\item \textbf{$\rho_a$ (proportion of the population willing to be vaccinated)}: As of summer 2022, 78\% of the total population of the U.S. had received at least one dose of the COVID-19 vaccine, with booster shots and new vaccination rates tapering off \cite{KFF}.  As this is a population that had ample access to freely available vaccines we can assume that the value $0.78$ represents the proportion of the population willing to be vaccinated and use it as a base value of $\rho_a$ for both areas. A major study conducted in 2020, prior to the availability of a COVID-19 vaccine, indicated a higher vaccine willingness among low and middle income countries (LMICs) than among high income countries (HICs):  80.3\% of those surveyed in LMICs, compared to 65\% in HICs, expressed willingness to take a COVID-19 vaccine if one became available \cite{Nature}.  
However, vaccine willingness varies significantly by country \cite{Nature}.  
For sensitivity analysis, we consider a low value of $\rho_a = 0.50$, and a high value of $\rho_a = 1.00$.




\item \textbf{$\rho^V_a$ (initial proportion of the population vaccinated)}: We consider the early stage of vaccine rollout, in late 2020 and early 2021, when neither area had yet distributed any vaccine: $\rho^V_a = 0$.  


\item \textbf{$\rho_a^I$ (initial rate of new infections)}:  To estimate the initial rate of new infections in each area, we first estimate the prevalence of COVID-19 in late 2020, and then divide this prevalence by the average duration of infectiousness.  Due to the significance of asymptomatic cases in COVID-19 transmission, which limit direct estimation of prevalence, a variety of approaches have been used to estimate COVID-19 prevalence in England  \cite{Nature_Eng_prev}, the U.S. \cite{PLOS_US_prev, Science_US_prev}, and Canada \cite{JMIR_Canada_prev} at different periods of the pandemic.  
One study estimates a prevalence in the U.S. of $1.4\%$ on December 31, 2020 
\cite{PLOS_US_prev}.  Then, $\rho_a^I$ can be estimated as $0.014 r_a^d = 3.6 \times 10^{-3}$ for the baseline, with a range of $[2.6 \times 10^{-3}, 4.9 \times 10^{-3}]$.  






\item \textbf{$V_a(t)$ (Rate of vaccinations available at time $t$, people/day)}:  
$V_a(t)$ is an input parameter that reflects the vaccine allocation in each area. 



\item \textbf{$r_0,\Delta r_a, r^d_a$,  (Rate out of infectious state, proportion/day)}:  The rate out of the infectious state is given by $r^d_a = r_0 + \Delta r_a$, where $r_0$ is the baseline rate absent testing, and $\Delta r_a$ is the change in that rate due to testing.

Absent testing, we estimate a base value of 3.9 days in the infectious state with a range from 2.5 to 5.1 days, or $r_0 = \frac{1}{3.9}$ (range $[\frac{1}{5.1},\frac{1}{2.5}]$). This reflects contributions from both symptomatic infections and asymptomatic infections. Amongst symptomatic infections, we estimate a one to three day period (base value of two days) of infectiousness before isolation \cite{Byrne, AronnaEtAl_2021_Epidemics}.  Additionally, using an estimate of 23.5\% (range: $[0.17, 0.30]$) for the percentage of infections that are asymptomatic \cite{RasmussenPopescu} and an assumed 10-day period of infectiousness for asymptomatic cases \cite{CDC_durationisolation}, asymptomatic cases increase the average duration of infectiousness to 3.9 days.

For the donor country, we assume a random daily testing rate of 5\% of those in infectious states, as in \cite{AronnaEtAl_2021_Epidemics}, 100\% test effectiveness in those states,  0\% effectiveness in the exposed states, and immediate test results. 
 Under these conditions, the average time spent in the infectious state decreases to 3.4 days, with a range of 2.2 to 4.5 days.  Thus, we use a base value of $\Delta r_a = \frac{1}{3.4} - \frac{1}{3.9} = 0.035$  (range: $[0.026, 0.060]$) for the donor country.  We use $\Delta r_a = 0$ for the nondonor country.

\item \textbf{$\gamma_a$ (behavioral infection multiplier)}: $\gamma_a$ reflects the adjustment to the disease transmission rate that is affected by the geographical and behavioral characteristics of the area $a$, such as population density and cultural norms around social distancing.  The model of Volpert \textit{et al.} uses both a high-transmission infection rate and a low-transmission infection rate to model behavioral differences in two subpopulations of a same geographic area and find estimated behavioral infection multipliers ranging from 3.8 to 9.7 \cite{VolpertVitalySharma_EcoComp_2021}.  
However, across different geographic areas, each of which is comprised of subpopulations of varying transmission characterstics, we would not expect to see as stark disparity as they found.  Therefore, 
in the baseline scenario, we use $\gamma_a = 1$ for both the donor and nondonor countries, and we use $\gamma_{nondonor}=2$ for sensitivity analysis.

\end{itemize}

\subsection{Other Parameters of SEIR-V (Table \ref{tab:param2})}

Table \ref{tab:param2} lists parameters reflecting the epidemiological characteristics of COVID-19.  These parameters are estimated as follows.

\begin{itemize}
    \item \textbf{$r^I$ (rate of out of the exposed state, proportion/day)}: Lauer \textit{et al.} estimate the incubation period of COVID-19 to be approximately 5 days \cite{Lauer}, giving a rate from the exposed state E into the infectious state I of $r^I = \frac{1}{5}$.
    
    \item $p^D, p_V^D$ \textbf{(mortality rates for unvaccinated (resp., vaccinated) classes, proportion/day)}: The CDC reports a total of 6,812,040 and 2,866,517 COVID-19 cases among unvaccinated and vaccinated people, respectively, during the period April 4 - December 25, 2021 in 25 U.S. jurisdictions.  In that same period and in the same jurisdictions, the CDC reports 94,640 and 22,567 COVID-19-associated deaths of unvaccinated and vaccinated people, respectively \cite{CDC_mortality}.  Thus, we estimate an unvaccinated case 
    mortality rate of $p^D = 0.014$, and a vaccinated case 
    mortality rate of $p_V^D = 0.0079.$




   \item \textbf{$a_0, \Delta a$ (infection rates)}: 
   For the baseline infection rate of the initial strain of the virus, before considering differences between areas, we assume that each infectious person's interactions will lead to 0.6 potential exposures per day: 
   $a_0 = 0.6$.  $\Delta a$ is the increase in infection rate of the new variant above $a_0$ (see equation (2)).  When the Delta variant of COVID-19 emerged, it was found to have roughly double the infectiousness as the Alpha variant \cite{Hansen, Yale}.  Thus, we let $\Delta_a = 0.6$ (range: $[0.3, 0.9]$) to model a variant that has twice the infection rate as the initial strain.  
    



   \item $v^u$ \textbf{(upper limit on proportion infectious):} During the COVID-19 pandemic, social distancing and masking measures typically followed rising case rates.  To reflect this, we reduce the effective number of infectious individuals, $\mathcal{V}$, by multiplying by a factor that decreases linearly in the number infectious; see equation (5). The parameter $v^u$ is the proportion infectious in the population at which transmission stops, making it an upper limit on this proportion. Rather than estimate it directly, we tune $v_u$ and $a_0 + \Delta a$ so that in our baseline scenario a target percentage of the donor country population will be infected with the new variant.  It is estimated that about 30\% of Americans had been infected with COVID-19 prior to the emergence of the omicron variant, and that nearly 60\% of the population had been infected a few months later \cite{Wapo}. Thus, we tune these parameters so that approximately 30-40\% of the donor country population will be infected with the new variant; we find $v^u = 0.03$ (range: $[0.025, 0.10]$).   
    
    
    \item $p^e, p^r$ \textbf{(vaccine effectiveness):}   
    Using the case counts from the supplementary material of Eyre et al. \cite{rr}, we see that the positivity rate of contacts of partially vaccinated index cases is 0.60 times the positivity rate of contacts of unvaccinated index cases. Thus, we use $p^e = 0.6$ (range: $[0.5, 0.8]$).   %
    %
    Data from Eyre et al. \cite{rr} show that vaccinated contacts were 0.62 as likely to test positive than unvaccinated contacts.  Thus, we use $p^r = 0.6$ (range: $[0.5, 0.8]$).  
    
    \item \textbf{$n$ (person-days of infection before appearance of new variant):}  We select values of $n$ so that the variant emerges approximately one-third of the way through the 180-day simulation.  In our baseline scenario, this value is $n=45,000$.  As described in Section \ref{sec:results}, the \textit{large nondonor initial cases} scenario uses $n=60,000$, and the \textit{global variant} scenario uses either the baseline value or $n=90,000$.

    \item \textbf{$L$ (time lag for a new variant to spread between areas)}:  
    Based on estimates of emergence dates of the omicron variant in the Netherlands, the U.S., Canada, and Scotland, after the first case was detected in South Africa, we estimate $L = 15$ days \cite{whoomicron, wikiomicron, msn}.
    
    \item\textbf{$T_D$, $p$, $k$ (time for variant to dominate)}: 
   Based on reports of omicron dominance    from South Africa, Scotland, Canada, the U.S., and Korea, we estimate the time from emergence in an area to dominance to be $T_D = 25$ days \cite{whoomicron, wikiomicron, msn, koreatimes}. \textit{Emergence} is defined as when the variant represents $1\%$ of those currently infectious, so $p = 0.01$. Given $T_D$ and $p$, the infection rate for the mixture of the two variants is given in equation (2) and involves a decay rate parameter, $k=\ln[(1-p)/p]/T_D$.  
    

    \item \textbf{$B$ (daily budget of vaccine doses)}: To complete vaccinations in a reasonable time for the populations considered, we use a vaccination budget $B = 1500$ 
 (range: $[1000,2000]$) doses per day. 

    \item \textbf{$T$ (time horizon)}: The baseline scenario uses a time horizon of 180 days (approximately six months). Sensitivity analysis examines the impact of 180-day policies over a 270-day or 360-day time horizon (9-12 months).  This reflects current understanding that natural immunity of the COVID-19 virus following an infection lasts at least 10 months \cite{LancetNews}, and therefore is consistent with our model formulation that assumes people who have recovered from COVID-19 remain in the Recovered class for the duration of the simulation and cannot become reinfected.

\end{itemize}

\section{Emergence of Variant}
\label{sec:emerge}

Holleran \textit{et al.} use a stochastic model for the emergence of a more infectious variant \cite{HolleranEtAl_Opt}.  Here, we propose a simpler, deterministic version of their model for how the infection rate changes over time due to a new variant. Initially, all areas are assumed to have constant infection rate, reflecting an initial mix of variants (\cite{HolleranEtAl_Opt}):
\begin{equation} \label{eq:alpha_0}
    \alpha_a(t) = a_0 \gamma_a.
\end{equation}
Like them, we assume that the variant emerges only in nondonor areas, which are largely low-income countries with less healthcare access and higher vulnerability to long infections; this assumption is relaxed in Section \ref{sec: emergence}.  The variant has an infection rate that is larger by $\Delta a$.  

\begin{figure}[!ht]
    \centering
    \includegraphics[scale = 0.45]{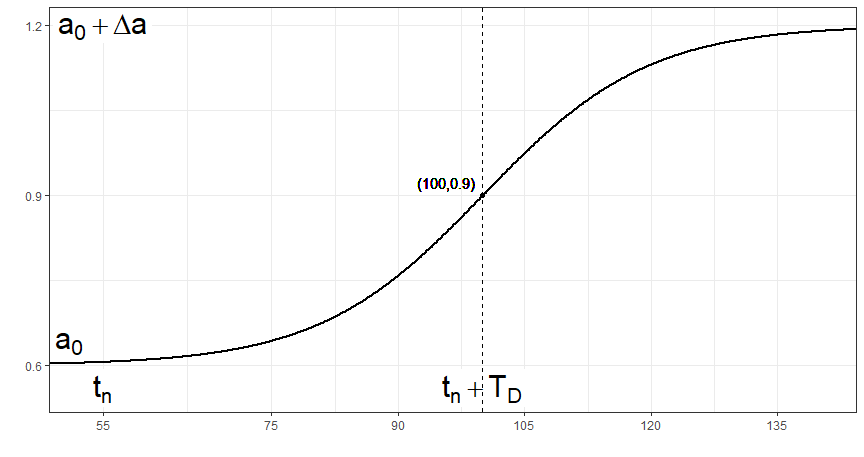}
    \caption{Equation (\ref{eq:alpha_app}) with $t_n = 55,\, T_D = 45,\, a_0 = 0.6,\, \Delta a = 0.6,\, p = 0.01$}
    \label{fig:alpha}
\end{figure}

In contrast to Holleran \textit{et al.}, we assume the new variant emerges once a \textit{fixed} number $n$ of unvaccinated infectious person-days have accumulated in the nondonor area. For a given scenario, policy, and $n$, let $t_n$ be the time when the new variant appears, $m$ the area where it appears, and $T_D$ the delay until it becomes dominant (half of infections). The infection rate for the mixture of the two variants in area $m$ is
\begin{equation}\label{eq:alpha_app}
    a(t) = a_0 + \frac{\Delta a}{1+e^{-k(t-(t_n + T_D))}}.
\end{equation}
 See Figure \ref{fig:alpha}. The rate parameter is $k = \ln[(1-p)/p]/T_D$. The new variant is assumed to reach other areas with a lag of $L$ days. 

The remainder of the derivation follows Holleran \textit{et al.} \cite{HolleranEtAl_Opt}. Including the behavior factor, the time-varying infection rate is (\cite{HolleranEtAl_Opt}): 
\begin{equation} \label{eq:alpha_a_app}
  \begin{aligned}
    \alpha_m(t) & = a(t) \gamma_m , \quad t \ge t_n  \\
    \alpha_a(t) & = \alpha_m(\, \max\{t-L,\,0\} \, )\gamma_a/\gamma_m \text{   for }a \neq m.
  \end{aligned}
\end{equation}
To simulate SEIR-V under any given vaccation policy $V_a(t)$, we discretize the system of differential equations in time steps of one day. The discretized system is solved iteratively over $t$ for all areas to find $t_n$ and the area $m$ where the variant emerges. The variant appears in day $t^*$,  which is the smallest integer for which (\cite{HolleranEtAl_Opt}):
\begin{equation}
\label{eq:t}
\sum_{a \in \mathcal{A \setminus \mathcal{D}}} \sum_{t=0}^{t^*}{I_a(t)} \ge n.
\end{equation}
The variant emerges in the nondonor area $m$ with the largest number of unvaccinated infection-days $\sum_{t=0}^{t^*}{I_a(t)}$. We interpolate between days, setting (\cite{HolleranEtAl_Opt}):
\begin{equation} \label{eq:tn}
t_n = t^* - 1 + 
      \frac{\sum_{a \in \mathcal{A} \setminus \mathcal{D}} \sum_{t=0}^{t^*}{I_a(t)} - n}
           {\sum_{a \in \mathcal{A} \setminus \mathcal{D}} {I_a(t^*)}}.
\end{equation}

\end{document}